\documentclass{elsart5p}
\usepackage{graphics}
\usepackage{graphicx}
\usepackage{epsfig}
\usepackage{amssymb}
\usepackage{amsmath}

\def\fmn#1#2{\mbox{${\textstyle \frac{#1}{#2}}$}}
\newcommand{\dpce}{\mbox{$dp\to \{pp\}_{s}n$}}
\newcommand{\dpcepol}{\mbox{$\pol{d}p\to \{pp\}_{s}n$}}

\newcommand{\dpcedSpol}{\mbox{$\pol{d}p\to \{pp\}_{s}\Delta^{0}$}}
\newcommand{\dpceX}{\mbox{$dp\to \{pp\}_{s}X$}}
\newcommand{\dpceXS}{\mbox{$dp\to \{pp\}_{s}X$}}
\newcommand{\dpcepolX}{\mbox{$\pol{d}p\to \{pp\}_sX$}}
\newcommand{\nppdelta}{\mbox{$np \to p\,\Delta^{0}$}}
\newcommand{\nppdeltapol}{\mbox{$\pol{n}p \to \pol{p\,}\Delta^{0}$}}

\newcommand{\ppdpi}{$pp \rightarrow d\pi^{+}$}

\begin{document}

\begin{frontmatter}

\title{Excitation of the $\Delta(1232)$ isobar in deuteron
charge exchange on hydrogen at 1.6, 1.8, and 2.3~GeV}

\author[hepi,ikp]{D.~Mchedlishvili},
\author[pnpi]{S.~Barsov},
\author[ipnup]{J.~Carbonell},
\author[hepi,ikp]{D.~Chiladze},
\author[jinr1,erlangen]{S.~Dymov},
\author[pnpi]{A.~Dzyuba},
\author[ikp]{R.~Engels},
\author[ikp]{R.~Gebel},
\author[jinr2]{V.~Glagolev},
\author[ikp,pnpi]{K.~Grigoryev},
\author[munster]{P.~Goslawski},
\author[ikp]{M.~Hartmann},
\author[jinr1]{O.~Imambekov},
\author[ikp]{A.~Kacharava\corauthref{cor1}},
\ead{a.kacharava@fz-juelich.de}\corauth[cor1]{Corresponding
author.}
\author[ikp]{V.~Kamerdzhiev},
\author[hepi,basel]{I.~Keshelashvili},
\author[munster]{A.~Khoukaz},
\author[jinr1]{V.~Komarov},
\author[krakow]{P.~Kulessa},
\author[jinr1]{A.~Kulikov},
\author[ikp]{A.~Lehrach},
\author[hepi]{N.~Lomidze},
\author[ikp]{B.~Lorentz},
\author[hepi,jinr1]{G.~Macharashvili},
\author[ikp]{R.~Maier},
\author[ikp,jinr1]{S.~Merzliakov},
\author[munster]{M.~Mielke},
\author[ikp,pnpi]{M.~Mikirtychyants},
\author[ikp,pnpi]{S.~Mikirtychyants},
\author[hepi]{M.~Nioradze},
\author[ikp]{H.~Ohm},
\author[munster]{M.~Papenbrock},
\author[ikp]{D.~Prasuhn},
\author[ikp]{F.~Rathmann},
\author[ikp]{V.~Serdyuk},
\author[ikp]{H.~Seyfarth},
\author[ikp]{H.J.~Stein},
\author[erlangen]{E.~Steffens},
\author[ikp]{H.~Stockhorst},
\author[ikp]{H.~Str\"oher},
\author[hepi]{M.~Tabidze},
\author[ikpros]{S.~Trusov},
\author[jinr1,msu]{Yu.~Uzikov},
\author[ikp,pnpi]{Yu.~Valdau},
\author[ucl]{C.~Wilkin}

\address[hepi]{High Energy Physics Institute, Tbilisi State University, GE-0186 Tbilisi, Georgia}
\address[ikp]{Institut f\"ur Kernphysik and J\"ulich Centre for Hadron Physics,
Forschungszentrum J\"ulich, D-52425 J\"ulich, Germany}
\address[pnpi]{High Energy Physics Department, Petersburg Nuclear Physics
Institute, RU-188350 Gatchina, Russia}
\address[ipnup]{Institut de Physique Nucl\'{e}aire, Universit\'e Paris-Sud, IN2P3-CNRS, F-91406 Orsay
Cedex, France}
\address[jinr1]{Laboratory of Nuclear Problems, JINR, RU-141980 Dubna, Russia}
\address[erlangen]{Physikalisches Institut II, Universit\"at Erlangen-N\"urnberg,
D-91058 Erlangen, Germany}
\address[jinr2]{Laboratory of High Energies, JINR, RU-141980 Dubna, Russia}
\address[munster]{Institut f\"ur Kernphysik, Universit\"at M\"unster, D-48149
M\"unster, Germany}
\address[basel]{Department of Physics, University of Basel, Klingelbergstrasse
82, CH-4056 Basel, Switzerland}
\address[krakow]{H.~Niewodnicza\'{n}ski Institute of Nuclear Physics PAN, PL-31342
Krak\'{o}w, Poland}
\address[ikpros]{Institut f\"ur Kern- und Hadronenphysik,
Forschungszentrum Rossendorf, D-01314 Dresden, Germany}
\address[msu]{Department of Physics, M.~V.~Lomonosov Moscow State University,
RU-119991 Moscow, Russia}
\address[ucl]{Physics and Astronomy Department, UCL, Gower Street, London, WC1E 6BT, UK}

\date{Received: \today / Revised version:}

\begin{abstract}
The charge-exchange break-up of polarised deuterons \dpcepol, where the final
$\{pp\}_s$ diproton system has a very low excitation energy and hence is
mainly in the $^{1\!}S_0$ state, is a powerful tool to probe the spin-flip
terms in the proton-neutron charge-exchange scattering. Recent measurements
with the ANKE spectrometer at the COSY storage ring at 1.6, 1.8, and 2.27~GeV
have extended these studies into the pion-production regime in order to
investigate the mechanism for the excitation of the $\Delta(1232)$ isobar in
the $\pol{d}p\to \{pp\}_{s}X$ reaction. Values of the differential cross
section and two deuteron tensor analysing powers, $A_{xx}$ and $A_{yy}$, have
been extracted in terms of the momentum transfer to the diproton or the
invariant mass $M_X$ of the unobserved system $X$. The unpolarised cross
section in the high $M_X$ region is well described in a model that includes
only direct excitation of the $\Delta$ isobar through undistorted one pion
exchange. However, the cross section is grossly underestimated for low $M_X$,
even when $\Delta$ excitation in the projectile deuteron is included in the
calculation. Furthermore, direct $\Delta$ production through one pion
exchange only reproduces the angular dependence of the difference between the
two tensor analysing powers.
\end{abstract}


\end{frontmatter}

It was pointed out many years ago that quasi-free ($p,n$) or ($n,p$)
reactions on the deuteron can, in suitable kinematic regions, act as a spin
filter that selects the spin-dependent contribution to the neutron-proton
elastic charge-exchange cross section~\cite{POM1951}. The comparison of this
reaction with free backward elastic scattering on a nucleon target provides
information on the neutron-proton backward elastic scattering amplitudes.
This field has been comprehensively surveyed in Ref.~\cite{LEH2010}.

Theory suggests that even more detailed information on the $np$
charge-exchange amplitudes could be obtained by measuring the charge-exchange
break-up of tensor polarised deuterons, \dpcepolX~\cite{BUG1987}. By
selecting two final protons with low excitation energy, typically
$E_{pp}<3$~MeV, the emerging diproton is dominantly in the $^{1\!}S_0$ state.
The reaction then involves a spin flip from the initial spin triplet of the
deuteron to the spin singlet of the $S$-wave diproton. In the well studied
neutron case, $X=n$, the amplitude in impulse approximation is proportional
to that in $np \to pn$, times a form factor that reflects the overlap of the
initial deuteron and final diproton wave functions. This approach describes
quantitatively a range of measurements of the differential cross section,
tensor and vector analysing powers, and spin correlation coefficients in the
\dpce\ reaction provided that the contamination of $P$- and higher partial
waves in the final $pp$ system is taken into account~\cite{CAR1991}.

The ANKE-COSY collaboration has carried out a series of experiments to deduce
the energy dependence of the spin-dependent $np$ elastic amplitudes by
identifying the neutron channel in the \dpcepol\ reaction~\cite{MCH2013}.
However, the same experimental data clearly show the possibility of extending
these studies into the pion-production regime in order to investigate the
excitation of the $\Delta(1232)$ isobar.

It was first demonstrated at SATURNE that the $\Delta(1232)$ can indeed be
produced in the \dpcedSpol\ charge-exchange reaction at a deuteron beam
energy $T_d=2.0$~GeV~\cite{ELL1987,ELL1989,SAM1991}. In analogy to the final
neutron case, it is expected that the highly inelastic deuteron
charge-exchange measurements correspond to a spin transfer from the initial
neutron to final proton in the \nppdeltapol\ process with a spectator proton.
This would give valuable information on the spin structure in the excitation
of the $\Delta$ isobar.

\begin{figure}[hbt]
\begin{center}
\includegraphics[width=0.43\columnwidth,clip]{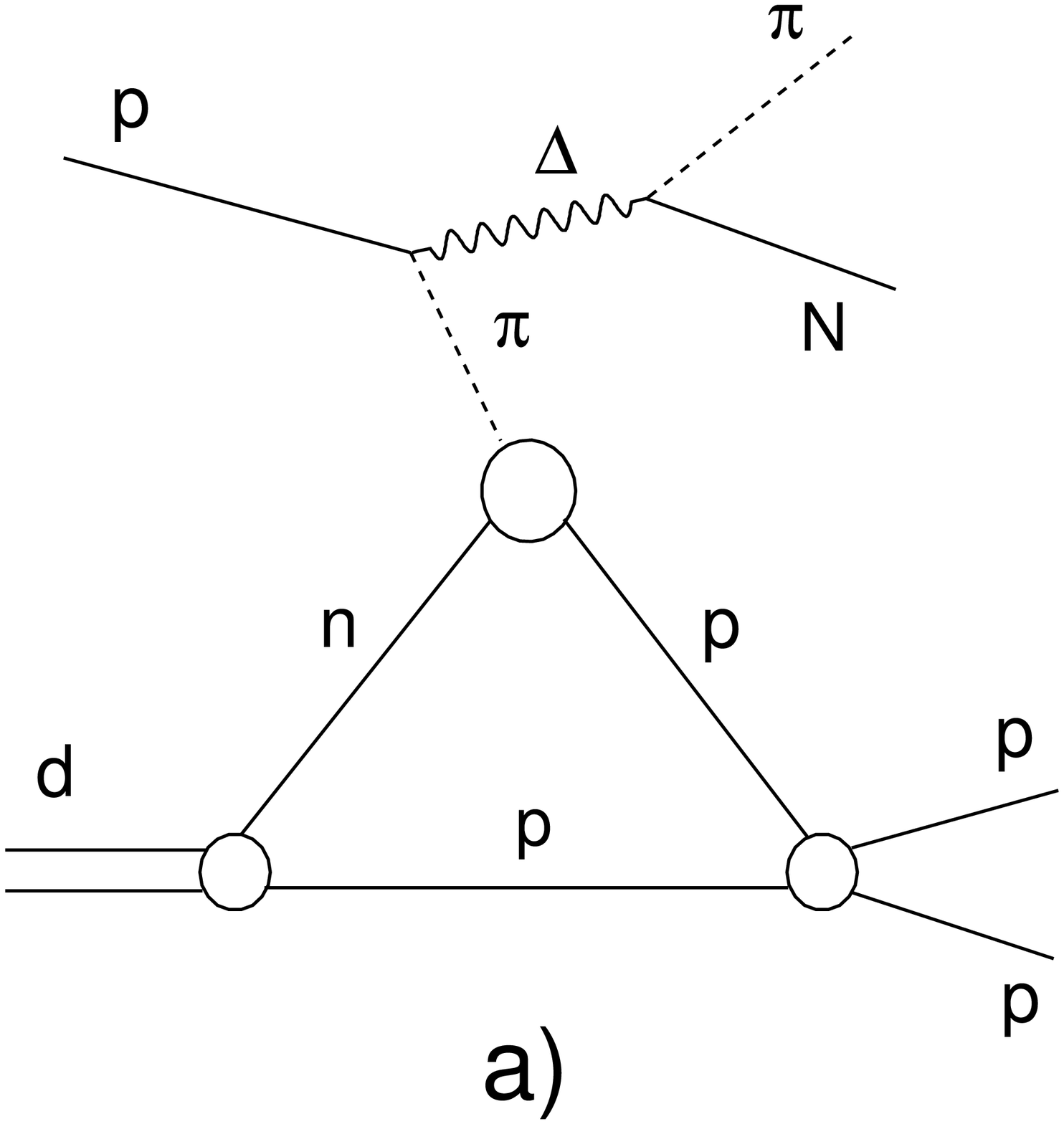}
\hspace{10mm}
\includegraphics[width=0.43\columnwidth]{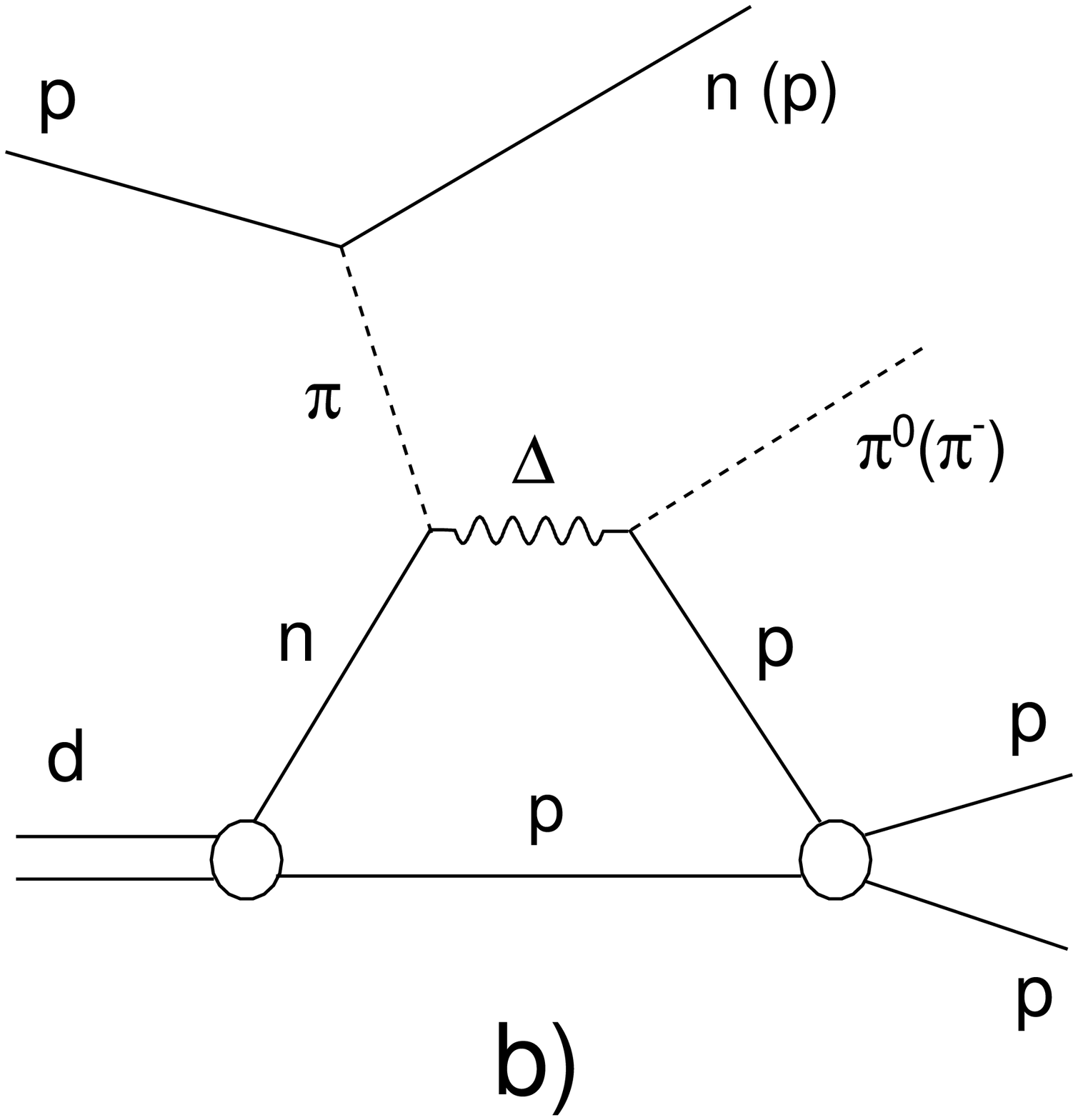}
\caption{\label{diag_delta} The one-pion-exchange contribution
to $\Delta(1232)$ production in the deuteron charge-exchange
break-up reaction. (a) The direct (D) term. (b) The exchange
(E) term.}
\end{center}
\end{figure}

The one-pion-exchange (OPE) model is quite successful in describing the
unpolarised cross section of the $pp\to \Delta^{++}n$ reaction as shown, for
example, in Ref.~\cite{DMI1986}. The model contains direct (D) and exchange
(E) terms which, when applied to the $dp\to \{pp\}_sN\pi$ reaction,
correspond to the diagrams in Figs.~\ref{diag_delta}a and b, respectively. It
should be noted that in impulse approximation the direct diagram contains the
same triangle loop as in the \dpce\ reaction, i.e., the same $d\to
\{pp\}_{s}$ form factors.

Measurements have been carried out at the COoler SYnchrotron
(COSY)~\cite{MAI1997} of the Forschungszentrum J\"ulich using transversely
polarised deuteron beams with energies $T_{d}$ = 1.2, 1.6, 1.8, and 2.27~GeV
incident on an unpolarised hydrogen cluster-jet target~\cite{KHO1999}. A
detailed description of the ANKE magnetic spectrometer~\cite{BAR2001} used
for the deuteron charge-exchange studies, as well as the procedure for
identifying the reaction, can be found in Refs.~\cite{MCH2013,spin2010}.

In addition to displaying a well-separated neutron peak, the experimental
\dpceX\ missing-mass spectra also show a lot of strength at higher $M_X$ that
must be associated with pion production. The analysis of these data for both
the differential cross section and deuteron tensor analysing powers is
similar to that for $\dpcepol$, whose results have already been reported in
some detail~\cite{MCH2013}. Greater emphasis in this letter will therefore be
given to the results rather than the procedures.

Values of the absolute luminosity are required in order to extract normalised
cross sections. In this experiment this is achieved by measuring the
quasi-free $np \to d\pi^0$ reaction in parallel. The cross section for
producing this final state is smaller than that for \ppdpi\ by an isospin
factor of two. There are extensive measurements of the latter reaction and
these have been included in the amplitude analysis of the SAID
group~\cite{ARN1993}. An additional advantage of using quasi-free pion
production for normalisation is that the effect of the shadowing in the
deuteron largely cancels out between the \dpceX\ and $dp\to p_{\rm
spec}d\pi^0$ reactions, where $p_{\rm spec}$ is a spectator proton.
Furthermore, in both cases two fast hadrons have to be detected so that there
is less influence from any acceptance uncertainties.

The count rates of the $dp\to p_{\rm spec}d\pi^0$ reaction were corrected for
the track reconstruction and proportional chamber efficiencies and the dead
time of the data acquisition system. Monte Carlo simulations were performed
at all energies to take into account the effects of the ANKE acceptance on
the experimental data. More details on the luminosity determination are to be
found in Ref.~\cite{MCH2013}.

Monte Carlo simulations were performed also for the \dpceX\ reaction at all
three energies, in order to evaluate the ANKE acceptance. Events were
generated according to the simple one-pion exchange mechanism of
Fig.~\ref{diag_delta}a. By dividing the numbers of reconstructed events by
the total, two-dimensional acceptance maps were obtained in $\theta_{pp}$ and
$M_X$. The maximum value of the diproton polar angle $\theta_{pp}$ changes
slightly with energy and $M_X$, reaching $4.5^{\circ}$ in the laboratory
system. In order to avoid potentially unsafe regions, where the acceptance
drops very rapidly, the cut $\theta_{pp}<3^{\circ}$ was applied at all
energies, in both the simulation and analysis. The strong $S$-wave
final-state interaction (FSI) between the two measured protons was taken into
account according to the Migdal-Watson approach~\cite{WAT1952,MIG1953}, using
the $pp$ $^{1\!}S_0$ scattering amplitude~\cite{MOR1968}.

\begin{figure}[h]
\begin{center}
\includegraphics[width=0.9\columnwidth, angle=0]{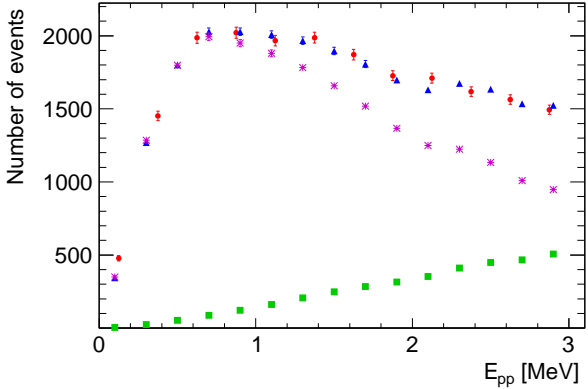}
\caption{\label{fig_epp} Experimental (red dots) and simulated $E_{pp}$
distributions, summed over all three beam energies. The simulation of the
$S$-wave contribution (magenta stars) includes a Migdal-Watson
factor~\cite{WAT1952,MIG1953}. The fitted value of the non-interacting
$P$-wave (green squares) corresponds to a total contribution of 15\% over this
$E_{pp}$ range. The overall simulation is shown by blue triangles. }
\end{center}
\end{figure}

Since direct production of the $\Delta$ isobar necessarily involves
relatively high momentum transfers, the $P$-wave contribution is
non-negligible. This effect is clearly observed in the comparison between the
uncorrected experimental and simulated $E_{pp}$ distributions for all events
shown in Fig.~\ref{fig_epp}. The $S$-wave term falls well below the data for
$E_{pp}>1$~MeV. Any FSI will be much weaker in the $P$-waves, so that the
weight for this contribution is proportional to the square of the $pp$
relative momentum, i.e., the diproton excitation energy. By fitting the two
terms together, it was found that for all beam energies the $P$-wave
contribution is about $15\%$ of the total event rate for $0<E_{pp}<3$~MeV.

\begin{figure}[h]
\begin{center}
\includegraphics[width=0.95\columnwidth, angle=0]{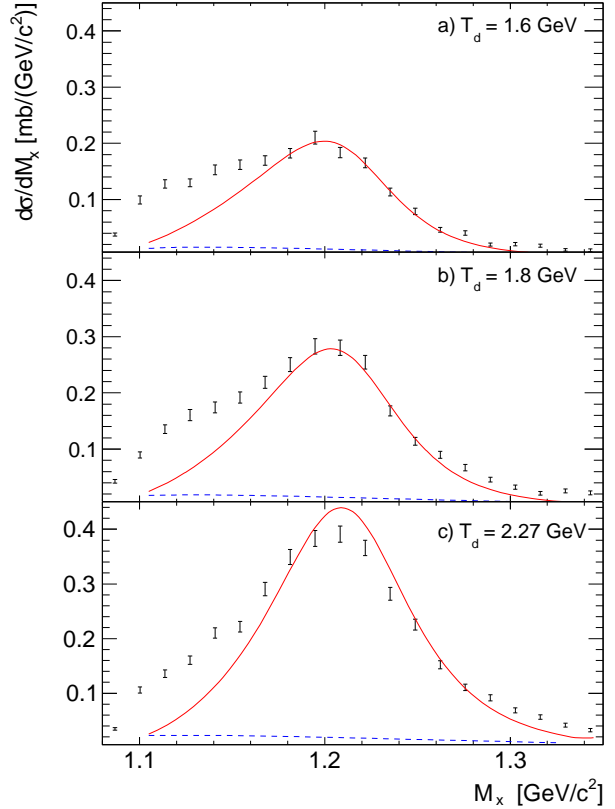}
\caption{\label{dsdm} Unpolarised differential cross section for the $\dpceX$
reaction with $E_{pp}<3$~MeV for $M_X>M_N+M_{\pi}$ at three deuteron beam
energies. The data are summed over the interval $0 < \theta_{\rm lab} <
3^{\circ}$ in the diproton laboratory polar angle. Only statistical errors
are shown; the overall normalisation uncertainties are less than 7\%. The
solid (red) curves correspond to the one-pion-exchange predictions for the
direct mechanism of Fig.~\ref{diag_delta}a. The dashed (blue) lines show the
contribution of the exchange mechanism (E) of Fig.~\ref{diag_delta}b.}
\end{center}
\end{figure}

The resulting \dpceX\ missing-mass cross sections are shown in
Fig.~\ref{dsdm} for $E_{pp}<3$~MeV. In the mass range accessible at COSY,
single pion production is dominated by the formation and decay of the
$\Delta(1232)$ isobar. It is therefore reassuring that the spectra at all
three beam energies are maximal for $M_X\approx 1.2$~GeV/$c^2$. The
evaluation of the isobar contribution to such spectra must depend on a
theoretical model, to which we now turn.

Expressions for the two-dimensional cross section for the direct $\Delta$
production are given explicitly in Ref.~\cite{DMI1986}. The resulting
one-dimensional cross section can be written as:
\begin{eqnarray}
\label{dsdmx}
\frac{d\sigma}{dM_X}=\frac {1}{128\pi^3m K^2s}\int_{t_1}^{t_2}\!dt\!
\int_0^{k_{\rm max}}\!k^2dk\,\rho(M_X)\,\overline{|\mathcal{M}_{fi}|^2},
\end{eqnarray}
where $\mathcal{M}_{fi}$ is the $dp\to\{pp\}_s\Delta^0$ transition matrix
element and $M_X$ is the $\Delta^0$ invariant mass. Here $m$ is the proton
mass, $K$ the incident c.m.\ momentum, $s$ the square of the c.m.\ energy,
and $k$ the internal momentum in the final diproton. In the ANKE experiment
the maximum value $k_{\rm max}$ is fixed by the cut in the excitation energy
$E_{pp}=k^2/m<3$~MeV. The integration over the four-momentum transfer $t$ in
Eq.~(\ref{dsdmx}) corresponds to the interval in the diproton polar angle in
the laboratory system $0^\circ <\theta_{\rm lab}<3^\circ$.

The spectral function $\rho(M_X)$ in Eq.~(\ref{dsdmx}), which accounts for
the finite width of the $\Delta$-isobar, has the form~\cite{ELL1985}
\begin{equation}
\label{rhof}
\rho(M_X)=\frac{1}{\pi}\frac{M_\Delta\Gamma \,Z(M_X^2,t)}
{(M_X^2-M_\Delta^2)^2+\Gamma^2M_\Delta^2}
\end{equation}
with
\begin{equation}
\label{zf}
Z(M_X^2,t)=\frac{p^2(M_\Delta^2,t)+\kappa^2}{p^2(M_X^2,t)+\kappa^2},
\end{equation}
where $p^2(M_X^2,t)=\lambda(M_X^2,m^2,t)/4M_X^2$ and
$\lambda(a,b,c)$ is the triangle function. The width of the
$\Delta$-isobar is
\begin{equation}
\label{gamma}
\Gamma=\Gamma_0\left(\frac{p(M_\Delta^2,m_\pi^2)}
{p(M_X^2,m_\pi^2)}\right)^{\!3} Z(M_X^2,m_\pi^2),
\end{equation}
with the following parameters: $M_\Delta=1.232$~GeV/$c^2$,
$\Gamma_0=0.115$~GeV/$c^2$, and $\kappa=0.180$~GeV/$c$.

If we consider only the $^{1\!}S_0$ final $pp$ state, the
spin-average square of the one-pion-exchange transition matrix
element in impulse approximation is
 \begin{eqnarray}
\label{mfi2}
&\overline{|\mathcal{M}_{fi}|^2}=
\frac{1}{6}\frac{[(M_X-m)^2-t][(M_X+m)^2-t]^2}{(t-m_\pi^2)^2\,3M_X^2}\times& \nonumber\\
&\left[\frac{f_{\pi }}{m_\pi}\frac{f^*}{m_\pi}8\sqrt{m^3}\sqrt{\frac{2}{3}}
 F_{\pi }^2(t)\sqrt{Z(M_X^2,t)} F(t,k^2) \right]^2q^2,&
\end{eqnarray}
where $\vec{q}$ is the total three-momentum transfer. The $d\to \{pp\}_s$
transition form factor $F(t,k^2)$, which includes the effects of the $S$- and
$D$-states in the deuteron~\cite{BUG1987}, has been evaluated using the CD
Bonn $NN$ interaction~\cite{MAC2000} for both the deuteron and the $pp$
$S$-wave scattering state. The $\pi NN$ and $\pi N\Delta$ coupling constants
used are $f_\pi=1.0$ and $f^*=2.15$, respectively, and a form factor
$F_{\pi}(t)=(\Lambda_{\pi}^2-m_\pi^2)/(\Lambda_{\pi}^2-t)$ has been
introduced at both vertices.

The size of the cut-off parameter $\Lambda_\pi$ in the form factors affects
the absolute values of the cross section predictions shown in
Fig.~\ref{dsdm}. The value chosen here, $\Lambda_\pi=0.5$~GeV/$c$, gives a
good description of the magnitude of the cross section at high $M_X$ for all
three beam energies. Its value is less than the $\Lambda_\pi=0.63$~GeV/$c$
found from a one-pion-exchange fit to inclusive $pp\to \Delta^{++} n$
results~\cite{DMI1986}, but it agrees well with $\Lambda_\pi=0.5$~GeV/$c$
obtained from exclusive $pp\to pn\pi^+$ data~\cite{IMA1988}.

The simple direct one-pion-exchange model for the \nppdelta\ amplitude
describes well the data in Fig.~\ref{dsdm} at high $M_X$, though it must be
stressed that this calculation neglects the 15\% $P$-state contribution shown
in Fig.~\ref{fig_epp}. However, the approach underestimates enormously the
low mass results. This failure must be more general than the specific
implementation of the model because the $\Delta$ is a $p$-wave pion-nucleon
resonance. There can therefore be little strength at low $M_X$ and this
suggests  that one should search for other mechanisms that might dominate
near the $\pi N$ threshold.

Exactly the same problem for the cross section in our angular domain was
noted in the pioneering SATURNE experiment~\cite{ELL1985}, where one pion
exchange was only successful at high $M_X$. To investigate this further, the
authors compared the small-angle hydrogen target data, $p(d,{pp})X$, with
quasi-free production in deuterium, $d(d,{pp})X$. From this it is clear that
the excess of events at low $M_X$ is mainly to be associated with isospin
$I=\frac{1}{2}$ $\pi N$ pairs rather than the $I=\frac{3}{2}$ of direct
$\Delta$ production~\cite{SAM1991}.

It is easy to exclude the culprit being $s$-wave $N^*$ resonance
contributions to direct production. To get a rough estimate of the possible
effects, the $p$-wave one-pion-exchange model predictions were modified as:
\begin{equation}
 \left(\frac{d\sigma}{dm}\right)_{\!\!s}\approx\left(\frac{d\sigma}{dm}\right)_{\!\!p}
 \times\frac{2\sigma(S_{11})+\sigma(S_{31})}{\sigma(P_{33})}\times\frac{p_0^2}{p^2},
\label{S_contr}
\end{equation}
where $\sigma(S_{11})$, $\sigma(S_{31})$, and $\sigma(P_{33})$ are the $\pi
N$ elastic cross sections in the three partial waves noted, and $p_0$ and $p$
are the momenta of the final and intermediate pion, respectively. Such an
estimate indicates only a very tiny extra strength at low $M_X$ and this
would have to be increased by several orders of magnitude in order to agree
with the experimental data. One must therefore seek an alternative
explanation to direct isobar production to describe the data.

There are some similarities between the \dpceX\ reaction and the inclusive
$dp\to dX$~\cite{BAN1973} or $\alpha p\to\alpha X$~\cite{MOR1992}
measurements that were dedicated to the search for the excitation of the
$N^*(1440)$ Roper resonance. Due to conservation laws, the isospin of the
unobserved state $X$ in these cases must be $I=\frac{1}{2}$ but this does not
necessarily correspond to an $N^*$ resonance. In fact the largest strength in
the data is seen at very low values of $M_X$, with only a small enhancement
arising from the $N^*(1440)$. The dominant effect is believed to be
associated with the excitation of the $\Delta(1232)$ isobar inside the
projectile deuteron or $\alpha$-particle~\cite{BAL1982,FER1995}. Although the
mechanism is driven by the $\Delta(1232)$, the pion and nucleon that make up
the state $X$ are produced at different vertices and so $X$ is not required
to be in a $p$-wave and to have isospin $I=\fmn{3}{2}$.

The exchange diagram E for the \dpceXS\ reaction is shown in
Fig.~\ref{diag_delta}b. However, a preliminary investigation of this
mechanism reported in Fig.~\ref{dsdm} suggests that, although it can provide
strength at low $M_X$, the overall magnitude is still far too small to
provide an adequate description of the data in this region~\cite{UZI2013}.
The relative reduction compared to the $dp\to dX$ or $\alpha p\to\alpha X$
calculations~\cite{BAL1982,FER1995} arises primarily from the spin-flip that
is inherent in the $d \to \{pp\}_s$ transition. We therefore turn to the
measurement of the tensor analysing powers for more clues.

In order to minimise systematic errors, several configurations of the
deuterium polarised ion source (with different vector and tensor components)
were employed. It is then necessary to determine the polarisations of each of
these deuteron beams from the scattering asymmetries in suitable nuclear
reactions with known analysing powers~\cite{CHI2006}. This was already
carried out for the analysis of the \dpcepol\ data~\cite{MCH2013}.

In the data analysis, we define the $z$-axis to lie along the beam direction
and the $y$-axis, which is along the upward normal to the COSY plane, is also
the stable spin axis. The $x$-axis is then defined by
$\hat{\vec{x}}=\hat{\vec{y}}\times\hat{\vec{z}}$. The three-momentum transfer
can be usefully split into longitudinal $q_z$ and transverse parts
$\vec{q}_t$, so that in general $\vec{q}=(q_t\cos\phi,q_t\sin\phi,q_z)$,
where we have introduced the azimuthal angle $\phi$ with respect to the
$x$-direction. The longitudinal component of the momentum transfer may be
written in terms of $q_t$ and the missing mass $M_X$.

When only the tensor polarisation is considered, the numbers
$N(q_t,M_X,\phi)$ of diprotons detected as a function of $q_t$, $M_X$, and
$\phi$ are given in terms of the beam polarisation $P_{zz}$ by\footnote{The
$z$ subscript here refers conventionally to the axis in the source frame. In
the COSY frame this becomes the $y$-axis.}
\begin{eqnarray}
\nonumber \frac{N(q_t,M_X,\phi)}{N_{0}(q_t,M_X)} &=& C_n\left\{1+\fmn{1}{2}P_{zz}\!
\left[A_{xx}(q_t,M_X)\sin^2\phi\right.\right.\\
&&\left.\left. \hspace{18mm}+A_{yy}(q_t,M_X)\cos^2\phi\right]\right\}\!.
\label{eq:ce_1}
\end{eqnarray}
The value of $C_n$, the luminosity of the polarised relative to the
unpolarised beam, was determined through measurements of single fast
spectator protons~\cite{MCH2013}. The same form of Eq.~(\ref{eq:ce_1}) can be
used at the calibration energy 1.2~GeV to determine the beam polarisation and
at the higher energies of 1.6, $1.8$, and $2.27$~GeV to extract the tensor
analysing powers of the \dpcepol\ and \dpcepolX\ reactions. Details on the
count-rate calibration and the procedure for the beam polarisation
determination are to be found in Ref.~\cite{spin2010}.

Due to limited statistics, it was not possible to measure $A_{xx}$ and
$A_{yy}$ as functions of two variables. Data were binned instead in either
$M_x$ or in $q_t$, summing over the full range of the other variable. Since
the acceptance is also a function of two variables, in such a procedure the
acceptance will influence the measurements of the analysing powers. In order
to minimise such effects in the analysis, the polarised and unpolarised data
were weighted with the inverse of the two-dimensional acceptance that was
evaluated for the extraction of the unpolarised cross section.

\begin{figure}[hbt]
\begin{center}
\includegraphics[width=0.8\columnwidth]{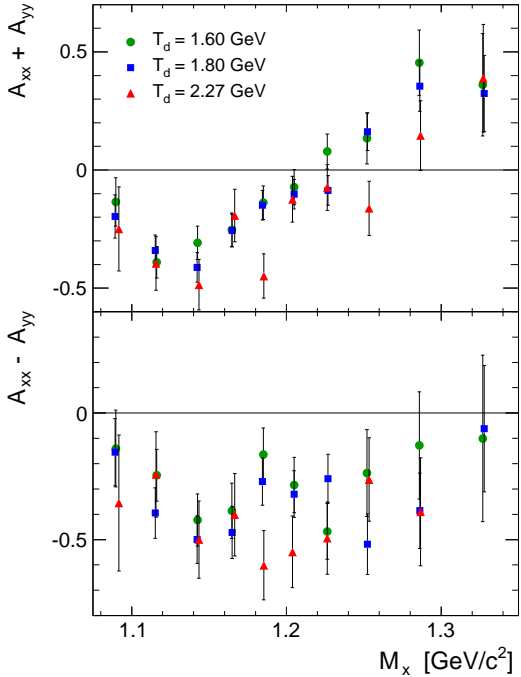}
\caption{\label{diffsum} The sum and difference of the Cartesian tensor
analysing powers for the \dpcepolX\ reaction with $E_{pp}<3$~MeV at three
different beam energies. The data are corrected for the detector acceptance
and summed over the range $0^{\circ} < \theta_{\rm lab} < 3^{\circ}$ in
diproton laboratory polar angle. Though the error bars are dominantly
statistical, they include also the uncertainties from the beam polarisation
and relative luminosity $C_n$. In addition, there is an overall uncertainty
of up to 4\% due to the use of the polarisation export technique~\cite{spin2010}.}
\end{center}
\end{figure}

The possible existence of two mass regions, where different mechanisms might
dominate, is also reflected in the behaviour of the tensor analysing power
shown in Fig.~\ref{diffsum}. After summing the data over the momentum
transfer, the sum and difference of the deuteron Cartesian tensor analysing
powers $A_{xx}$ and $A_{yy}$ are presented as functions of the missing mass
$M_X$. [These combinations are proportional to the spherical tensor
components $T_{20}$ and $T_{22}$.] No significant changes in the results were
found when considering the stronger cut $E_{pp} < 2$~MeV, which might reduce
any dilution of the analysing power signals by the $P$-waves apparent in
Fig.~\ref{fig_epp}.

It is interesting to note that the minimum in $A_{xx}+A_{yy}$ is at
$M_X\approx 1.15$~GeV/$c^2$, which is precisely the region where there is the
biggest disagreement with the cross section predictions of Fig.~\ref{dsdm}.
Furthermore, the values of $A_{xx}+A_{yy}$ seem to be remarkably stable,
showing a behaviour that is independent of beam energy. Hence, whatever the
mechanism is that drives the reaction, it seems to be similar at all
energies. The error bars on $A_{xx}-A_{yy}$ are larger since in this case,
according to Eq.~(\ref{eq:ce_1}), the slope in $\cos 2\phi$ has to be
extracted from the data. As a consequence it is harder to draw as firm
conclusions on the analysing power differences.

The SPESIV spectrometer at SATURNE had high resolution but very small angular
acceptance. The $T_d=2$~GeV data were therefore taken at discrete values in
the laboratory diproton production angle, typically in steps of $\approx
2^{\circ}$. The limited acceptance also meant that only a linear combination
of $A_{xx}$ and $A_{yy}$ (the ``polarisation response'') could be determined
and it was only at the larger angles that this approached a pure $A_{yy}$
measurement. The ANKE data have been analysed in much finer angular bins but
the $M_X$ distribution in the $\Delta$ region of the cross section and
polarisation response is quite similar to that measured at SATURNE at
$\theta_{\rm lab} = 2.1^{\circ}$~\cite{ELL1987,SAM1991}, which is in the
middle of the ANKE angular range.

\begin{figure}[hbt]
\begin{center}
\includegraphics[width=0.9\columnwidth]{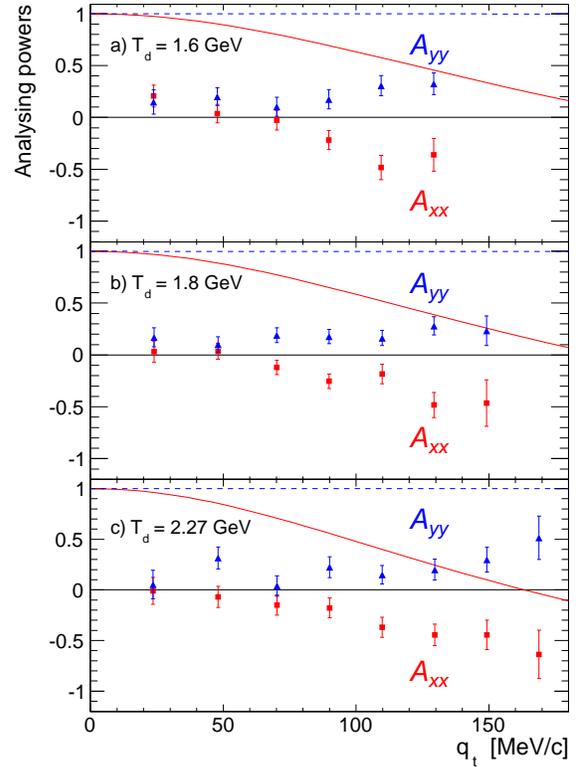}
\caption{\label{AnaPow} Acceptance-corrected tensor analysing powers $A_{xx}$
and $A_{yy}$ of the \dpcepolX\ reaction with $E_{pp}<3$~MeV at three deuteron
beam energies as a function of the transverse momentum transfer $q_t$. Only
high mass data ($1.19<M_X<1.35$~GeV/$c^2$) are considered. Note that in the
forward direction, $q_t=0$ and $A_{xx}=A_{yy}$. Though the error bars are
dominantly statistical, they include also the uncertainties from the beam
polarisation and relative luminosity $C_n$. In addition, there is an overall
uncertainty of up to 4\% due to the use of the polarisation export technique.
The one-pion-exchange predictions are shown by the blue dashed line for
$A_{yy}$ and red solid for $A_{xx}$.}
\end{center}
\end{figure}

Although the direct $\Delta$ production model of Fig.~\ref{diag_delta}a fails
to describe the differential cross section data of Fig.~\ref{dsdm} near the
pion production threshold, the situation is much more satisfactory at high
$M_X$. To investigate this region further, the data have been summed over the
range $1.19<M_X<1.35$~GeV/$c^2$ and the tensor analysing powers $A_{xx}$ and
$A_{yy}$ evaluated as functions of the transverse momentum transfer $q_t$.
The results at the three energies are shown in Fig.~\ref{AnaPow}.

Within the experimental uncertainties, the values of both $A_{xx}$ and
$A_{yy}$ at fixed $q_t$ seem to be largely independent of the beam energy.
This is consistent with a similar feature found for the data at fixed $M_X$
shown in Fig.~\ref{diffsum}. This suggests that there is a common reaction
mechanism at all three energies. Another important point to note is that the
signs of $A_{xx}$ and $A_{yy}$ are opposite to those measured in the
\dpcepol\ reaction~\cite{MCH2013} though, unlike the neutron channel case,
they tend to be very small at $q_t\approx 0$.

Estimates for the $\pol{d}p\to \{pp\}_s\Delta^0$ analysing powers can be
easily made for the direct one-pion-exchange production amplitude of
Fig.~\ref{diag_delta}a. In the non-relativistic limit this gives
\begin{equation}
\label{mfi}
M_{fi} \sim  \vec{q_{\pi}}\cdot\vec{\Psi}^+ \varphi_p F(t,k^2)\,\vec{q}\cdot \vec{\varepsilon},
\end{equation}
where $\vec{\Psi}$ is the vector-spinor of the $\Delta$-isobar,
$\vec{q_{\pi}}$ is the three-momentum of the virtual pion in the
$\Delta$-isobar rest frame, $\varphi_p$ is the spinor of the initial proton,
and $\vec{\varepsilon}$ represents the polarisation vector of the deuteron.

It follows from the $\vec{q}\cdot \vec{\varepsilon}$ factor of
Eq.~(\ref{mfi}) that only deuterons with magnetic quantum number $M=0$, when
quantised along the direction of the three-momentum transfer $\vec{q}$, can
lead to $\Delta$ production in the one-pion-exchange model. Due to the
$\Delta$--$p$ mass difference, $q_z$ is non-zero in the forward direction and
$\vec{q}$ then lies along the beam direction, so that in this limit
$A_{xx}=A_{yy}=1$.

For an arbitrary production angle, the tensor analysing powers become
\begin{equation}
A_{xx}=1- 3{q_t^2}/q^2\ \textrm{and}\ A_{yy}=1.
\end{equation}
Since the longitudinal momentum transfer depends upon the mass distribution
of the $\Delta$, $A_{xx}$ can only be estimated by integrating numerically
over the spectral shape. However, as can be seen from Fig.~\ref{AnaPow},
neither of the resulting predictions agrees even qualitatively with the
experimental data, which show very small analysing powers for $q_t \approx
0$. Nevertheless, if one looks instead at the combination $A_{xx}-A_{yy}$ it
seems the one-pion-exchange model does give a plausible description of the
data. In particular it offers an explanation as to why $A_{yy}>A_{xx}$ for
$\Delta$ production.

\begin{figure}[hbt]
\begin{center}
\includegraphics[width=0.8\columnwidth]{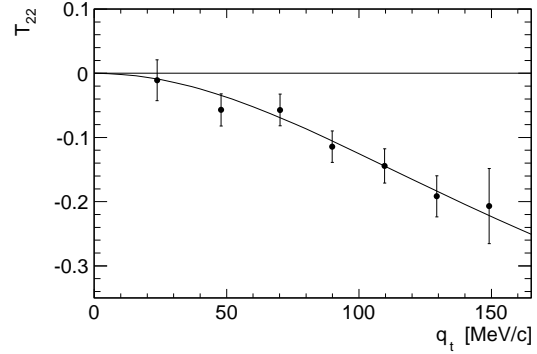}
\caption{\label{T22} Spherical tensor analysing power
$T_{22}=(A_{xx}-A_{yy})/2\sqrt{3}$ for the \dpcepolX\ reaction with
$E_{pp}<3$~MeV, averaged over the three beam energies studied. Though the
error bars are dominantly statistical, they include also the uncertainties
from the beam polarisation and relative luminosity $C_n$. In addition, there
is an overall uncertainty of up to 4\% due to the use of the polarisation
export technique. When the same approach is applied to the predictions of
the simple one-pion-exchange model of Fig.~\ref{diag_delta}a, the good
agreement shown by the curve is achieved.}
\end{center}
\end{figure}

To illustrate this in greater detail, we show in Fig.~\ref{T22} the simple
average over the three beam energies of the experimental data and the
one-pion-exchange predictions for the spherical analysing power
$T_{22}=(A_{xx}-A_{yy})/2\sqrt{3}$. Combining the energies in this way to
improve the statistics is reasonable because of the similarities shown by the
three sets of data in Fig.~\ref{AnaPow}.

It would be naive to expect that the rich features of the $pn\to\Delta^0n$
amplitude should be reproduced by considering only $\pi$ exchange. It is
important to note in this context that, in order to reproduce both $A_{xx}$
and $A_{yy}$ for the simpler \dpcepol\ reaction, all three spin-spin $np\to
pn$ amplitudes have to be used in the modeling of the
observables~\cite{BUG1987}. One may therefore hope that our data will provide
further impetus to the construction of more refined $pn\to\Delta^0n$ models.

Although we have not arrived at a satisfactory description of the low $M_X$
cross section or the high $M_X$ analysing powers, it may, nevertheless, be
helpful to consider the measurement of other observables in the \dpceX\
reaction. $\Delta$ production will be studied in the near future in inverse
kinematics with a polarised proton beam incident on a polarised deuterium gas
cell, $\pol{p}\pol{d}\to \{pp\}_sX$~\cite{CHI2012}, where the two slow
protons will be detected in silicon tracking telescopes~\cite{SCH2003}. This
will allow the studies reported here to be continued up to the maximum COSY
proton energy of $T_p\approx 2.9$~GeV.

A particularly intriguing possibility is the measurement in coincidence of a
proton or $\pi^-$ constituent from the state $X$. This would give access to
the tensor polarisation of $X$ which must, of course, vanish for an $s$-wave
pion-nucleon system. This would therefore provide yet another tool to
separate the $s$- and $p$-wave components of the system $X$.

In summary, we have measured the differential cross section and two tensor
analysing powers $A_{xx}$ and $A_{yy}$ in the highly inelastic deuteron
charge-exchange \dpcepolX\ reaction, where the effective mass $M_X$ of the
state $X$ indicates that pion production is involved. The high $M_X$ part of
the cross section data at all three beam energies studied is well reproduced
by a simple one-pion-exchange model. However, though this model fails to
reproduce the measured values of $A_{xx}$ and $A_{yy}$, the description of
$A_{yy}-A_{xx}$ as a function of $q_t$ shown in Fig.~\ref{T22} is
unexpectedly good and suggests that pion exchange does describe some of the
spin dependence in $\Delta$ production.

In addition to a possible direct $\Delta^0(1232)$ peak, there is a surprising
amount of production in the $s$-wave $\pi N$ region. Attempts to explain this
in terms of $\Delta$ excitation in the projectile deuteron give much too low
cross sections. Strength in this region could also arise from higher-order
diagrams involving a $\Delta N$ residual interaction~\cite{GAR1990}, which
have been neglected here. Although the other observables that will be
measured may cast more light on the reaction mechanism, further theoretical
work is needed in order that these data may be reliably related to the
$\pol{n}p \to \pol{p}\Delta^0$ reaction.

We are grateful to other members of the ANKE Collaboration for their help
with the experiment and to the COSY crew for providing such good working
conditions. This work has been supported by  the COSY FFE, the Shota
Rustaveli National Science Foundation, the Heisenberg-Landau programme, and
the European Union Seventh Framework Programme under grant agreement
n$^{\circ}$~283286.

%
%

\end{document}